\documentclass[11pt]{article}
\pdfoutput=1

\usepackage{amsmath,amssymb,comment,cite,esint,graphicx,mathtools}
\usepackage[margin=.8in,letterpaper]{geometry}
\usepackage[colorlinks=false]{hyperref}
\usepackage[affil-it]{authblk}
\usepackage{url}
\usepackage{color}

\numberwithin{equation}{section}
\let\originalleft\left
\let\originalright\right
\renewcommand{\left}{\mathopen{}\mathclose\bgroup\originalleft}
\renewcommand{\right}{\aftergroup\egroup\originalright}
\mathcode`\*="8000
{\catcode`\*=\active\gdef*{\mathclose{}\,\mathopen{}}}

\newcommand{\bean}{\begin{eqnarray}}
\newcommand{\eean}{\end{eqnarray}}
\newcommand{\eqs}[1]{Eqs. (\ref{#1})}
\newcommand{\eq}[1]{Eq. (\ref{#1})}
\newcommand{\meq}[1]{(\ref{#1})}
\newcommand{\fig}[1]{Fig. \ref{#1}}
\newcommand{\ppa}[2]{\left(\frac{\partial}{\partial #1}\right)^{#2}}

\newcommand{\ppn}[2]{\frac{\partial #1}{\partial #2}}

\newcommand{\hsp}{\hspace{0.1mm}}    %space

\newcommand{\bea}{\begin{eqnarray}}
\newcommand{\eea}{\end{eqnarray}}
\newcommand{\grad}{\nabla}

\newcommand{\eqn}{&=&}
\newcommand{\non}{\nonumber \\}

\begin{document}
\title{\textbf{Light rings and long-lived modes in quasiblack hole spacetimes}}

\author{
Minyong Guo, Zhen Zhong, Jinguang Wang, Sijie Gao$^{\ast}$}
\date{}

\maketitle

\vspace{-10mm}

\begin{center}
{\it Department of Physics, Beijing Normal University,
Beijing 100875, China\\\vspace{2mm}
}
\end{center}

\vspace{8mm}

\begin{abstract}
It has been argued that ultracompact objects, which possess light rings but no horizons, may be unstable against gravitational perturbations. To test this conjecture, we revisit the quasi-black hole solutions, a family of horizonless spacetimes whose limit is the extremal Reissner-Nordstr\"om black hole. We find a critical parameter at which the light rings just appear. We then calculate the quasinormal modes of the quasi-black holes. Both the WKB result and the numerical result show that long-lived modes survive for the range where light rings exist, indicating that horizonless spacetimes with light rings are unstable. Our work provides a strong and explicit example that light rings could be direct observational evidence for black holes.

\end{abstract}

\vfill{\footnotesize Email: minyongguo@bnu.edu.cn,\,zhenzhong@mail.bnu.edu.cn,\,jinguang@mail.bnu.edu.cn,\,sijie@bnu.edu.cn.\\$~~~~~~\ast$ Corresponding author.}

\maketitle

\newpage
\baselineskip 18pt
\section{Introduction}
Generally speaking, black holes are seen as the most fundamental particles in general relativity and modified theories of gravity. As the first black hole solution,  the Schwarzschild solution was found in 1916 \cite{Schwarzschild:1916}. Since then, theoretical properties of black holes have been extensively and deeply studied, including but not limited to the spacetime structure, thermodynamics, geodesics, quasinormal modes (QNMs) and the like. Compared to the great achievements in the theoretical aspect, the progress of experimental observations of black holes had been slow for a long time until the discoveries of gravitational waves made by LIGO and Virgo \cite{LIGOScientific:2016aoc, LIGOScientific:2016sjg, LIGOScientific:2017vwq} and the appearance of the first image of the black hole at the center of M87 photographed by the Event Horizon Telescope (EHT) \cite{EventHorizonTelescope:2019dse, EventHorizonTelescope:2019uob, EventHorizonTelescope:2019jan, EventHorizonTelescope:2021bee, EventHorizonTelescope:2021srq}.

The picture of M87* taken by EHT can be well explained by models of black hole \cite{EventHorizonTelescope:2019dse}. Furthermore, some parameters of M87* can be identified based on some specific black hole models \cite{Banerjee:2019nnj,Hou:2021okc, Kumar:2018ple, Hu:2020usx, Zhong:2021mty, Wang:2018prk,  Li:2020drn, Gan:2021xdl, Wang:2021ara, Zeng:2021dlj, Guo:2019lur,Yan:2019etp, Guo:2018kis,Zakharov:2011zz}. However, the existence of black holes has not been completely confirmed in terms of information encoded in the present photo of M87*. The main reason is  that it is hard to distinguish a black hole from an ultracompact objects (UCOs), which have no horizons. Previous works have found that  light rings (LRs), which are closed photon orbits, can exist not only outside  black holes, but also some UCOs\cite{Cardoso:2014sna, Guo:2020qwk,Cunha:2020azh, Ghosh:2021txu, Hod:2017zpi, Cunha:2017qtt}. Furthermore, it has been suggested that, apart from black holes, UCOs can also form shadow structures, like boson stars \cite{Cardoso:2014sna}, proca stars \cite{Herdeiro:2021lwl} and wormholes \cite{Wang:2020emr, Peng:2021osd}. In other words, it is going to get really tricky when black holes and UCOs can cast the same shadow structure. In addition, it has been found that UCOs can be relevant to other observations of the black holes at the centers of galaxies. For example, although most people believe there is no way to understand dark matter without involving black holes, many recent works revealed that some models related to UCOs are also able to explain known observations concerning dack matters including the rotational curve of stars\cite{Brito:2015yga, Narain:2006kx, Raidal:2018eoo, Cardoso:2019rvt}. 

A natural question then arises. Can we convince people that we are observing a black hole, other than a UCO, from the picture of M87* \cite{Cunha:2018gql,Lima:2021las,Cardoso:2019rvt, Zhang:2019glo}? First of all, as a kind of UCOs, wormholes are generally ruled out since to form a wormhole, exotic matters are always involved \cite{Morris:1988tu,Visser:1989kh,Poisson:1995sv,Wang:2015wmz}. Other typical UCOs include boson stars, that can be formed dynamically from a process of gravitational collapse and cooling \cite{Seidel:1993zk} and  Proca stars, that satisfy the Einstein equation and energy conditions\cite{Brito:2015pxa, DiGiovanni:2018bvo, Herdeiro:2021lwl}. Most people prefer to believe that the photos taken by EHT are indeed formed by black holes. Some researchers have speculated that although some existing UCOs can form shadow structures, such UCOs may have stability problems and they cannot exist long enough. For example, it has been argued that highly spinning horizonless UCOs with an ergoregion are unstable \cite{Friedman:1978, Chirenti:2008pf, Pani:2008bzt,Pani:2010jz}. For non-rotating black holes, in \cite{Keir:2014oka}, the author found a new mechanism suggesting that all ultracompact neutron stars with their radii $R< 3M$ might be unstable.

Along this line, Cardoso et al. have made a remarkable progress \cite{Cardoso:2014sna}. They focused on the spherically symmetric ultracompact stars. The radius of the star is always smaller than $3M$, and the outer spacetime is described by the Schwarzschild metric. Obviously, there exists an unstable LR at $r_{\text{LR}}= 3M$. Considering that the effective potentials of photons are divergent and positive at the center of the star, they showed that a stable LR has to be existent between the origin and the unstable LR in the radial direction. Furthermore, they investigated the QNMs of gravitational perturbations by focusing on specific constant-density stars and thin-shell gravastars with $2M<R<3M$, and showed the existence of very long-lived modes localized near the stable light ring, which may indicate such ultracompact stars are nonlinearly unstable under fragmentation. This result is very important and significantly supports that ultracompact stars may not be black hole mimickers \cite{Lemos:2008cv}.

However, the ultracompact stars discussed in \cite{Cardoso:2014sna} always have a radius smaller than $3M$, and thus the existence of LR is guaranteed. To investigate the relationship between the LR and the stability of the QNM, we need a one-parameter family of solutions where the LR appears at certain critical value,  and then we can check if this is also a critical point for the stability of the spacetime. Such solutions are not easy to find. Fortunately, a series of horizonless spacetimes, which are called quasi-black holes (QBH), have been constructed and discussed by Lue, Weinberg, Lemos, et.al. \cite{Lue:2000qr,   Lue:2000nm, Lemos:2003gx, Lemos:2007yh, Lemos:2009uk,Lemos:2010te}. QBH was motivated by the question whether static and  horizonless spacetimes can come arbitrarily close to a black hole. A charged dust model was constructed, which satisfies the Einstein-Maxwell equations \cite{Lemos:2003gx}. The make the solution approach  a black hole, it turns out that the dust must be extremal, i.e., the energy density of the dust must be equal to its charge density. This is a family of solutions parameterized by $c$. When $c=0$, it is just the extremal Reissner-Nordstr\"om (RN) black hole. This model can show how horizonless spacetimes continuously transfer to a true black hole. So when $c\to 0$, we expect the existence of LRs since black holes always have LRs. Also, we expect that the LRs disappear for some larger values of $c$ which correspond to configurations far away from the black hole. In this paper, we show that it is indeed the case.
 By using null geodesic equations, we find  a critical parameter $c=\sqrt{\frac{2}{27}}$. For $0<c<\sqrt{\frac{2}{27}}$, there always exist two LRs. For $c>\sqrt{\frac{2}{27}}$, the LRs disappear. This allows us to check the relation between the LR and the stability of the spacetime by calculating the quasinormal modes for QBHs. We find that the long-live modes survive for the range where LRs exist, indicating the instability of the spacetime. For the parameter range where the LRs do not exist, the long-live modes also disappear. Therefore, we use  QBHs to show explicitly that the existence of LR is closely related to the stability of spacetimes.

The remaining parts of the paper are organized as follows. In section 2, we give a quick review of the quasi-black holes. In section 3, we present a detailed study of LRs in quasi-black hole spacetimes.  In section 4, we study the quasinormal modes of gravitational perturbations for quasi-black holes. We summarize and discuss our results in section 5.

\section{Review on quasi-black holes}
The Einstein-Maxwell equation for charged dust takes the form \cite{Lemos:2003gx}
\bean
G_{ab}=8\pi(T_{ab}^{dust}+T^{em}_{ab})\,,
\eean
where
\bean
T_{ab}^{dust}=\rho u_au_b
\eean
with $\rho$ being the energy density and $u^a$ the four-velocity of the dust. The electromagnetic stress-energy tensor is given by
\bean
T^{em}_{ab}=\frac{1}{4\pi}\left(F_a \hsp^c F_{bc}-\frac{1}{4}g_{ab}F^{cd}F_{cd} \right)\,,
\eean
where the electromagnetic field strength $F_{ab}$ satisfies
\bean
\grad_bF^{ab}=4\pi \rho_e u^a
\eean
with $\rho_e$ being the charge density. We are interested in the extremal dust solution, i.e., $\rho=\rho_e$. It turns out that such solutions take the form \cite{Lemos:2003gx}
\bean
ds^2=-\frac{dt^2}{U^2}+U^2[dR^2+R^2(d\theta^2+\sin^2\theta d \phi^2)] \,.\label{dsR}
\eean
 By using Einstein's equation, one can show that
\bean
\frac{1}{R^2}\ppn{}{R}\left(R^2\ppn{U}{R}\right)=-4\pi U^3\rho \,.
\eean
We are interested in a series of solutions which can smoothly transfer from non-black hole solutions to black hole solutions. Such solutions can be obtained by choosing \cite{Lemos:2003gx}
\bean
U(R)=1+\frac{q}{\sqrt{R^2+c^2}}\,,
\eean
where $q$ is the total charge of the spacetime.

We see that when  $c\to 0$, the solution reduces to the extremal RN black hole and $R=0$ is the black hole horizon. In order to facilitate the following calculations, we introduce a non-negative parameter $z$ such tha
\bea
z^2=R^2+c^2\,.
\eea
Then the areal radius $r$ is related to $R$ by
\bean
r=RU=R+\frac{qR}{\sqrt{R^2+c^2}}=R+\frac{qR}{z} \,, \label{rru}
\eean
\begin{comment}
and the two metric components \eq{dsr} and \meq{dsR}  are related by
\bean
B\eqn\frac{1}{U^2}=\frac{z^2}{(z+q)^2} \,,\\\label{defW}
\frac{1}{\sqrt{A}}&\equiv&\frac{1}{W}=1+\frac{R}{U}\frac{dU}{dR}=\frac{z^2(q+z)}
{z^2(q+z)-qR^2} \,.
\eean
\end{comment}
From the components of the metric we can see that, when $c\neq0$ quasi-black hole spacetime could describe a spherically symmetric compact star with $R=0$ ($r=0$) being its center. While for $c=0$, the solution is just an extremal RN black hole and
\bean
r=R+q \,,
\eean
and the horizon $r=q$  corresponds to $R=0$ ($z=0$).

\section{Light rings}\label{lrs}
Unlike black holes, the existence of LRs for horizonless spacetimes are not guaranteed. In this section, we shall study photon orbits for QBHs and see whether LRs could exist. Since the QBH spacetimes are spherically symmetric, it is sufficient to focus the photon orbits on the
equatorial plane $\theta=\pi/2$. Then, the four-momentum of the photon takes the form
\bean
p^a=\dot t\ppa{t}{a}+\dot R\ppa{r}{a}+\dot\phi\ppa{\phi}{a} \,,
\eean
where the dot denotes the derivative of the affine parameter $\tau$. Considering the Killing vectors of the spacetime, the conserved energy and angular momentum are given by
\bean
E\eqn -g_{ab}p^a\ppa{t}{b}=-g_{tt}\dot t=\frac{1}{U^2}\dot t \label{ee} \,,\\
L\eqn g_{ab}p^a\ppa{\phi}{b}=g_{\phi\phi}\dot\phi=U^2R^2\sin^2\theta\dot\phi \label{ll} \,.
\eean
In addition, we  have the null condition
\bean
0=g_{ab}p^ap^b  \label{pp} \,.
\eean
By solving \eqs{ee},\meq{ll} and \meq{pp}, we get the radial equation
\bean
\dot R^2+\frac{L^2}{R^2U^4}=E^2 \,.
\eean
Next, we define the potential
\bean
V(R)=\frac{L^2}{R^2U^4} \label{vbigr} \,,
\eean
and then
\bean
V(R)=\frac{L^2}{R^2U^4}=\frac{L^2}{R^2\left(1+\frac{q}{\sqrt{c^2+R^2}}\right)^4}=
\frac{L^2z^4}{(z^2-c^2)(z+q)^4} \,.
\eean
The light rings occur at $V'(R)=0$ and $E^2=V(R)$.  Noting that $\frac{dz}{dR}=\frac{R}{z}$, we can use $V-E^2=\frac{dV}{dz}=0$ to determine the positions and the impact parameters of light rings instead. Thus, from $\frac{dV}{dz}=0$, we have
\bea\label{vp}
z^3-qz^2+2qc^2=0 \,.
\eea
For  $c=0$, we see immediately that there are two solutions $z=0$ and $z=q$, or $R=0$ and $R=q$. $R=q$ is just the light ring located outside the black hole horizon. By calculating the second derivative of the potential, we see that the LR is unstable in the radial direction. One may think that $R=0$ (or $r=q$) is also a LR. However, since the horizon $R=0$ is a null hypersurface, the only null geodesic on it is in the radial direction with no angular component. Thus, such a null geodesic does not form a closed orbit in space.

% except for the horizon $z=0$, the solution takes $z=q$ which corresponds to $R=q$ %and $r=2q$. In addition, considering %$\frac{d^2V(R=q)}{d}\propto\frac{d^2V(z=q)}{dz^2}=-\frac{1}{16q^2}<0$, we conclude %the light ring at $r=2q$ is unstable along the radial direction.

\begin{figure}[t]
  \centering
  \includegraphics[width=8cm]{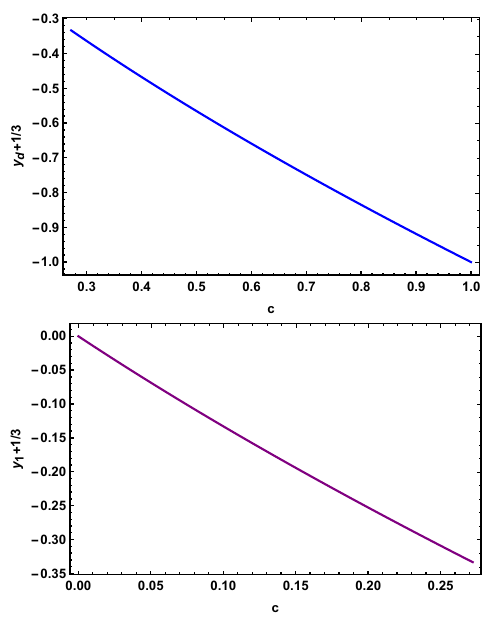}
  \caption{The functions $z_d=y_d+1/3$ and $z_1=y_1+1/3$. The upper panel corresponds to the real root of the cubic equation for $K<c^2<1$.
  The lower panel corresponds to the root $y_1$ when $0<c^2<K$.  They are both
  monotonic decreasing functions of $c$ and remain negative. Thus, these solutions do not give light rings. }
  \label{yd1}
\end{figure}

Next, we turn to the case $c\neq0$. To solve \eq{vp}, we let
\bean\label{rbc}
y=z-q/3>-q/3 \,.
\eean
In the following, for simplicity but without loss of generality, we set $q=1$. Then \eq{vp} becomes
\bean\label{seq}
y^3-\frac{1}{3}y+2c^2-\frac{2}{27}=0 \,.
\eean
In order to find the roots of the cubic equation, it's covenient to define
\bea
u=-\frac{1}{3}\,,\quad v=2c^2-\frac{2}{27}\,, \quad \text{and}\quad  \Delta=\left(\frac{v}{2}\right)^2+\left(\frac{u}{3}\right)^3
=c^2\left(c^2-\frac{2}{27}\right) \,.
\eea
Then, from $\Delta=0$, we can identify a critical constant $K\equiv\frac{2}{27}$. And, when $c^2>K$,  \eq{seq} has only one real root taking in this form
\bea
y_d=\frac{-1-\sqrt{3}i+i(i+\sqrt{3})\left(1-27c^2+3\sqrt{-6c^2+81c^4}\right)
^{2/3}}{6\left(1-27c^2+3\sqrt{-6c^2+81c^4}\right)^{1/3}} \,.
\eea
where $i$ is the imaginary unit. One can then easily check that $y_d+1/3<0$ is always true for $c^2>K$ which does not satisfy \eq{rbc} (see the upper panel  in Fig. \ref{yd1}), meaning  there is no light ring.

Next, we consider the case $\Delta=0$, that is, $c^2=K$. We find the roots of the cubic equation read
\bea
y_{m1}=y_{m2}=1/3\quad \text{and} \quad y_{m3}=-2/3 \,.
\eea
Obviously, $y_{m3}$ should be dropped and we find there are two degenerate light rings.

Now, let us turn to the case $\Delta<0$, that is $c^2<K$. We have three different real roots,
\bea
y_w&=&\frac{2}{3}\cos\frac{\Theta+2w\pi}{3}
\eea
where $w=-1, 0, 1$ and
\bea
\Theta=\arccos(1-27c^2).
\eea
From the lower panel of Fig. \ref{yd1}, we can see $(y_1+1/3)^2-c^2<0$ for $0<c^2<K$ and thus $y_1$ should be excluded. The other two LRs occur at
\bea
r_1\eqn \sqrt{z_0^2-c^2}\left(1+\frac{1}{z_0}\right) \,, \\
r_2\eqn \sqrt{z_{-1}^2-c^2}\left(1+\frac{1}{z_{-1}}\right) \,,
\eea
where
\bea
z_w=\frac{1}{3}\left(1+2\cos\frac{\Theta+2w\pi}{3}\right) \,.
\eea
with $w=-1, 0$. One can check that and the outer LR ($r=r_1$) is unstable and the inner LR ($r=r_2$)  is stable. This is consistent to the general conclusion for UCOs\cite{Guo:2020qwk,Cunha:2017qtt}.

Now we pay special attention to the regime $c\to 0$, i.e., the black hole limit. We can expand the roots to the  order of $c^2$ and find
\bea
R_0\eqn 1-\frac{5}{2}c^2, \\
R_{-1}\eqn c  \,, \label{brm}
\eea
which correspond to
\bea
r_0\eqn 2q-3\frac{c^2}{q}\,, \\
 r_{-1}\eqn \frac{q}{\sqrt{2}}+\frac{3}{2}c-\frac{9c^2}{4\sqrt{2}q} \label{srm} \,,
\eea
where we have put the charge $q$ back to the formula. We see that as $c\to 0$, $r_0\to 2q$ which just reduces to the light  ring of the extremal RN black hole. However, as $c\to 0$, $r_{-1}\to q /\sqrt{2}$. This result does approach the RN limit $c=0$, where such a light ring does not exist. To understand this apparent inconsistency, we notice that  \eq{srm} is obtained by substituting \eq{brm} into \eq{rru} and then taking the limit $c\to 0$. For the extremal RN solution, we let $c=0$ in \eq{rru} and then taking the limit $R\to 0$. This lead to $r=q$, which is not a LR as we have discussed.

\section{Long-lived QNM modes of a quasi-black hole spacetime}
In  section \ref{lrs}, we have found when $0<c^2<K$, the quasi-black hole can be seen as an UCO with an inner stable light ring and an outer unstable light ring. In this section, we are going to verify whether QBHs have stability problem under linear gravitational perturbations. More precisely, we would calculate the frequencies of QNMs to see if there are long-lived modes in the parameter range where the LRs exist. Considering that QNMs have been studied widely in the standard coordinates $(t,r,\theta,\phi)$, our strategy for dealing with the QNMs is that we first obtain the QNM equations in the coordinates $(t,r,\theta,\phi)$ based on a general formula for any spherically symmetric metric in \cite{Cardoso:2014sna}, and then we transform the equations into a form of the coordinate $R$ since it's more convenient to use $R$ for the calculations about QBHs as we have done in section \ref{lrs}. Now, let us rewrite the spherically symmetric metric in the form
\bean
ds^2=-B(r)dt^2+A(r)dr^2+r^2(d\theta^2+\sin^2\theta d\phi^2) \label{dsr}
\eean
It is not difficult to find that the two metric components in \eqs{dsr} and \meq{dsR}  are related by
\bean
B\eqn\frac{1}{U^2}=\frac{z^2}{(z+q)^2} \,,\\\label{defW}
\frac{1}{\sqrt{A}}&\equiv&\frac{1}{W}=1+\frac{R}{U}\frac{dU}{dR}=\frac{z^2(q+z)}
{z^2(q+z)-qR^2} \,.
\eean

To calculate the quasi-normal modes, we start with the master equation\cite{Cardoso:2014sna}
\bea\label{master}
\left[\frac{\partial^2}{\partial_t^2}-\frac{\partial^2}{\partial_{r_\ast}^2}
+V_{sl}(r)\right]\Psi(r,t)=0 \,,
\eea
which can describe perturbations of different fields in the background of the metric \meq{dsr}. In the potential $V_{sl}(r)$, when $l\ge s$, $s=0, 1, 2$ correspond to the perturbations of massless scalar fields, Maxwell fields and a generically anisotropic fluid, respectively.  The tortoise coordinate $r_\ast$ is defined by $dr/dr_\ast=\sqrt{B/A}$, and the potential $V_{sl}(r)$ takes the form
\bea
V_{sl}(r)=B\left[\frac{l(l+1)}{r^2}+\frac{1-s^2}{2rA}\left(\frac{B^\prime}{B}-
\frac{A^\prime}{A}\right)+8\pi(p_{\text{rad}}-\rho)\delta_{s2})\right]\,.  \label{vslsr}
\eea
Alternatively, from the relationship between $r$ and $R$, that is $r=RU$, $V_{sl}(r)$ can be reexpressed in the coordinate $R$ as
\bea
\bar{V}_{sl}(R)\equiv V_{sl}(r)=\frac{1}{U^2}\left[\frac{l(l+1)}{R^2U^2}-\frac{1-s^2}{RUW^2}\frac{d}{dR}\left(\log U+\log W\right)\frac{dR}{dr}+8\pi(p_{\text{rad}}-\rho)\delta_{s2}\right].\label{Vsl}
\eea
where $W\equiv\sqrt{A}$ is introduced in \eq{defW} And the the radial pressure $p_{\text{rad}}$ and the energy density $\rho$ of the QBH are needed to be introduced and we find they have the following expressions 
\bea\label{pex}
p_{\text{rad}}=T^r_r=-\frac{q^2R^2}{8\pi z^2(q+z)^4}\quad \text{and}\quad \rho=-T^t_t=\frac{q[qR^2+6c^2(1+z)]}{8\pi z^2(q+z)^4} \,,
\eea
By the way, we want to stress that by calculating
\bea
p_{\text{rad}}+\rho=\frac{3c^2(1+z)}{4\pi z^2(1+z)^4}>0 \,,
\eea
we can confirm that the weak energy condition always holds  for the QBH. Next, assuming a time dependence $\Psi(r, t)=\psi(r)e^{-i\omega t}$, from the \eq{master} we can see that the radial function $\psi(r)$ satisfies a Schr$\ddot{\text{o}}$dinger-like equation
\bea
\frac{d^2\psi}{dr_\ast^2}+[\omega^2-V_{sl}(r)]\psi=0,
\eea
with
\bea
r_\ast=\int_0^RUR\frac{dr}{dR}dR=R+\frac{\arctan(R/c)}{c}
+2\operatorname{arctanh}\frac{R}{\sqrt{c^2+R^2}} \,.
\eea
Alternatively, we have
\bea\label{req}
\frac{(c^2+R^2)^2}{(1+\sqrt{c^2+R^2})^4}\bar{\psi}''(R)+
\frac{2R(c^2+R^2)}{(1+\sqrt{c^2+R^2})^5}\bar{\psi}'(R)+
[\omega^2-\bar{V}_{sl}(R)]\bar{\psi}(R)=0\,,
\eea
where we have used $\bar{\psi}(R)\equiv\psi(r)$.

\begin{figure}[t]
  \centering
  \includegraphics[width=15cm]{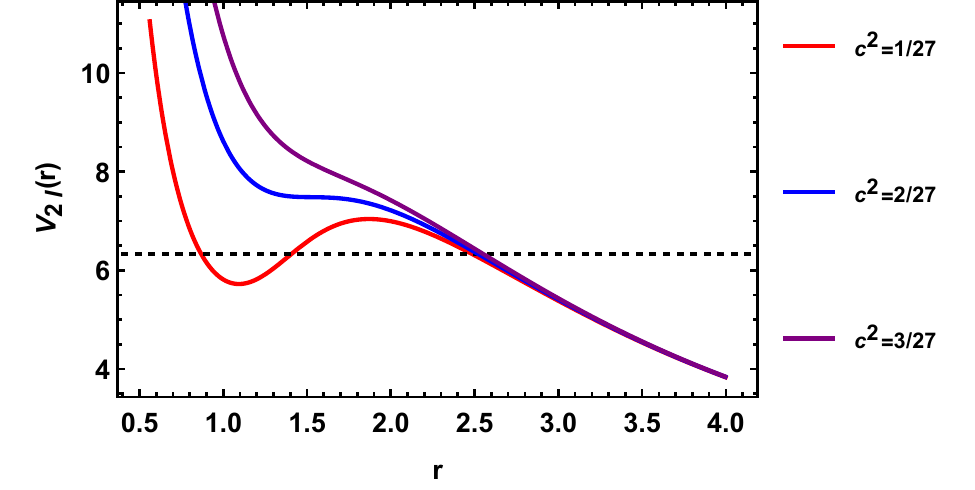}
  \caption{The variations of the effective potential function $V_{2l}$ with respect to $r$ for $c^2=1/27, 2/27, 3/27$, respectively. Here we choose $l=10\gg1$.}
  \label{vplot}
\end{figure}

\begin{figure}[t]
  \centering
  \includegraphics[width=12cm]{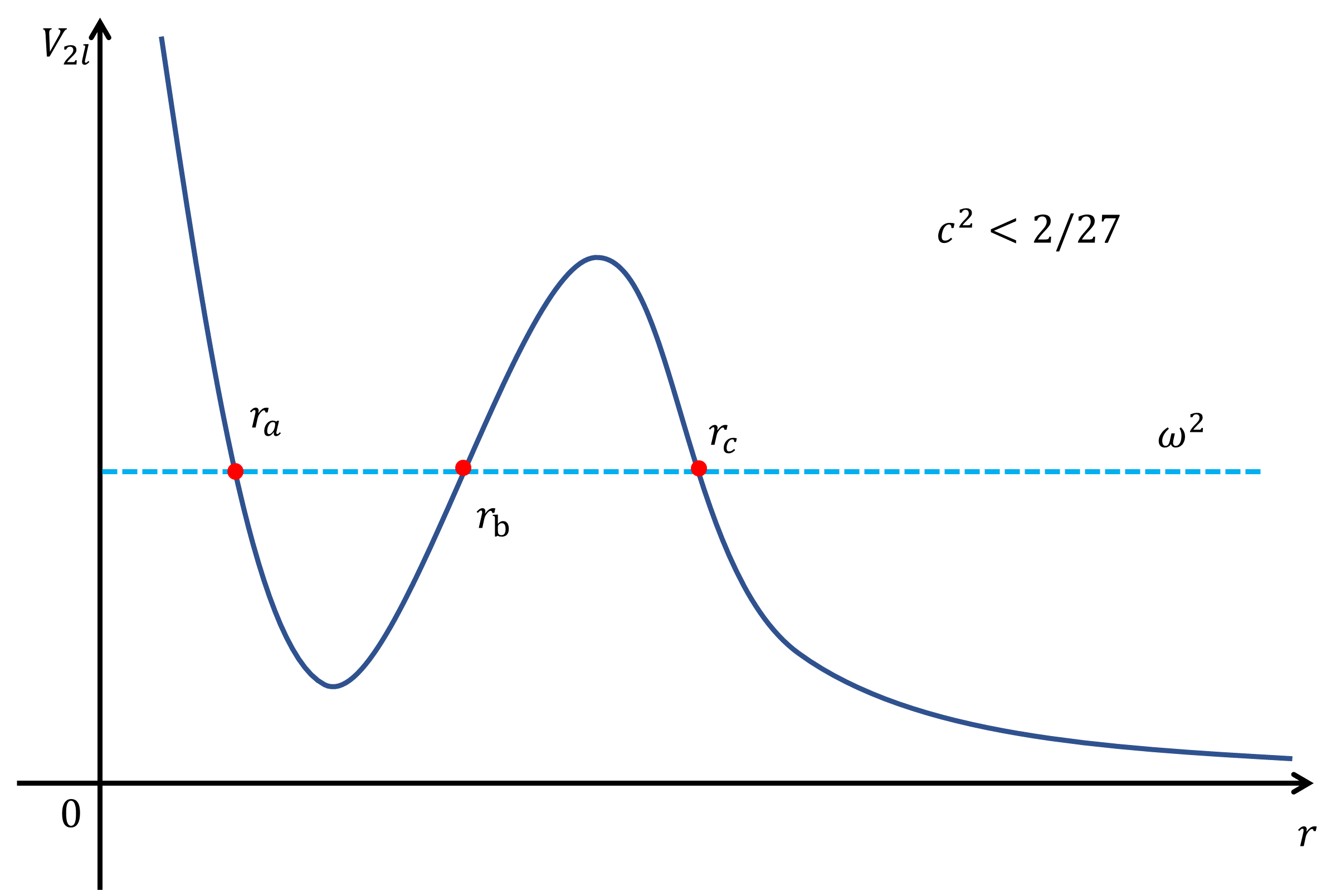}
  \caption{A diagram of the potential function $V_{2l}$ with respect to $r$ for $0<c^2<2/27$.}
  \label{diav}
\end{figure}

\begin{figure}[t]
  \centering
  \includegraphics[width=10cm]{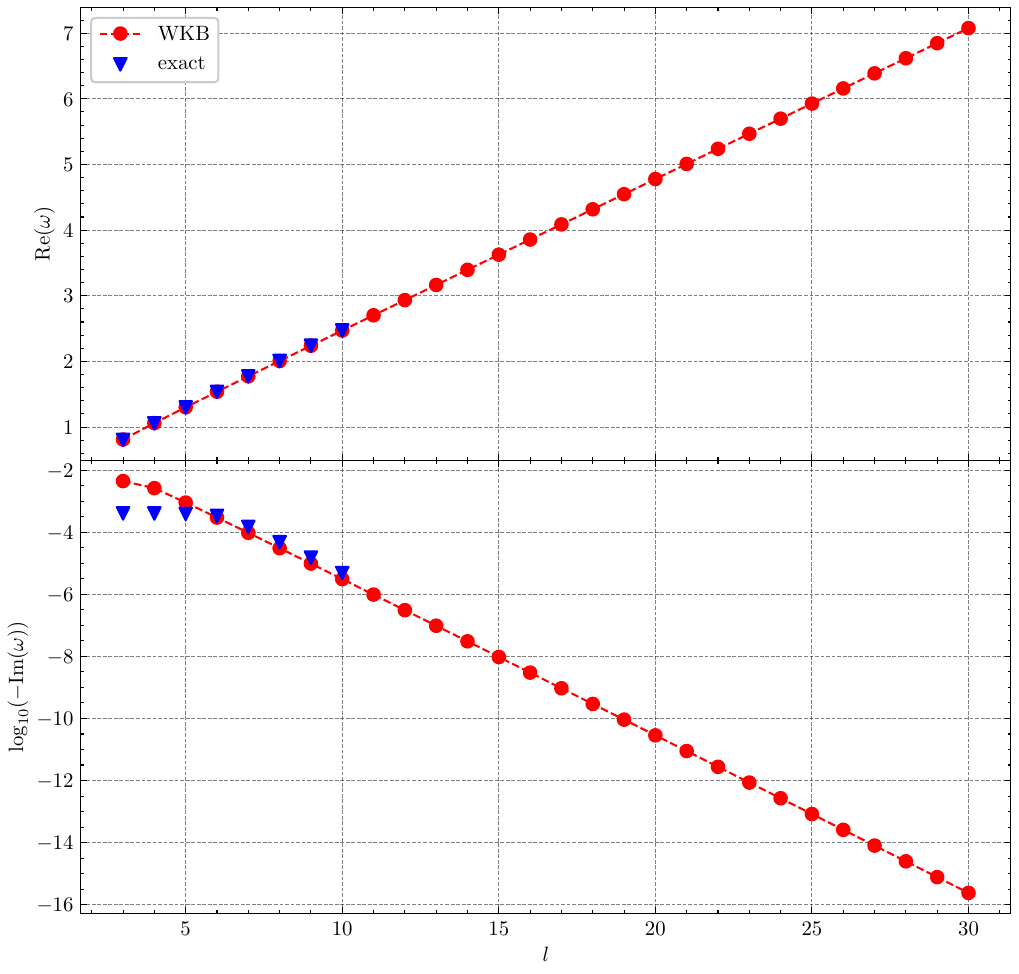}
  \caption{ Real and imaginary parts of the QNMs of the QBH with $c^2=1/27$. The red lines are the WKB results and the blue inverted triangles reveal the numerical results obtained from the direct integration method. From the lower panel, we can see that QNMs are indeed long-lived modes, which indicates a possible nonlinear instability of the spacetime.}
  \label{qnmw}
\end{figure}

In this article, we focus on  gravitational perturbations, that is, $s=2$. Thus at the center of the quasi-black hole, $R\to0$, we find
\bea
 r_\ast\to \frac{(\sqrt{c^2}+1)^2}{c^2}R\to0, \quad \bar{V}_{2l}\to \frac{c^4l(l+1)}{(1+\sqrt{c^2})^4R^2}=\frac{l(l+1)}{r_\ast^2}\,,
\eea
and at infinity, $R\to\infty$, we have
\bea
 r_\ast\to R, \quad \bar{V}_{2l}\to\frac{l(l+1)}{R^2}=\frac{l(l+1)}{r_\ast}\,.
\eea
Furthermore, we have at the center
\begin{comment}
\bea
r_\ast\to0, \quad \psi\sim C_1 r_\ast^{l+1}+C_2 r_\ast^{-l} \,,
\eea
or
\end{comment}
\bea\label{Rcon1}
R\to0, \quad \bar{\psi}\sim C_1 R^{l+1}+C_2 R^{-l} \,,
\eea
and at infinity
\begin{comment}
\bea
r_\ast\to\infty, \quad \psi\sim D_1 e^{-i\omega r_\ast}+D_2 e^{i\omega r_\ast} \,,
\eea
or
\end{comment}
\bea\label{Rcon2}
R\to\infty,\quad \bar{\psi}\sim D_1 e^{-i\omega R}+D_2 e^{i\omega R} \,.
\eea
Regular gravitational perturbations should have $C_2=0$ at the center, and at infinity, the gravitational perturbations should be outgoing, that is, $D_1=0$. In the following, we shall determine the values of $\omega$ in terms of the coordinate $R$. We rewrite $\bar{\psi}(R)$ in this form
\bea
\bar{\psi}(R)=R^{l+1}e^{i\omega R}\chi(R)
\eea
This asymptotic solution corresponds to an outgoing boundary condition at  infinity and a regular boundary condition near the center of the quasi-black black. The frequencies of the perturbations can be seen as compositions of quasinormal modes . With the specific boundary conditions  (\ref{Rcon1}) and (\ref{Rcon2}), the radial equation \meq{req} can be solved as an eigenvalue problem. Note that only some discrete eigenfrequencies $\omega$ can satisfy both the radial equation and boundary conditions. Since the frequency $\omega$ would be a complex number in general, one can always write the eigenfrequency as the form $\omega=\omega_R+i \omega_I$. As we assume $\Psi(r, t)=\psi(r)e^{-i\omega t}$, the amplitude of perturbation will grow exponentially when the imaginary part $\omega_I>0$,  which implies that the black hole is unstable at least at the linear perturbation level. Then in principle, we can identify the eigenvalues of $\omega$ by solving this equation for any $l\ge s$ numerically.

In practice, the behaviors of $\omega$ are very sensitive to the effective potential $V_{2l}$. In particular, we are interested in the eikonal regime, that is, $l\gg1$. In \fig{vplot}, we show examples of the effective potential $V_{2l}$ at $l=10$. Obviously, we can see that at $c^2=1/27$, the effective potential $V_{2l}$ has a local maximun and a local minimum, while for $c^2=3/27$, the effective potential $V_{2l}$ only has a local maximun. At $c^2=2/27$, the effective potential has a degenerate extreme point. It has been argued in \cite{Cardoso:2014sna,Keir:2014oka} that long-lived modes may be possible in the eikonal limit, that is, for $l\gg1$, compared to the usual modes, there may be some long-lived modes whose damping time grows exponentially with $l$, when the potential necessarily has a local minimum. In addition, from \eq{vslsr} we observe that when $l\gg\frac{r}{2A}\left(\frac{B^\prime}{B}-\frac{A^\prime}{A}\right)$ and $l\gg 8\pi r^2(p_{\text{rad}}-\rho)$, the effective potential $V_{sl}\simeq\frac{l^2}{r^2U^2}$, which has the same expression as the effective potential $V=\frac{L^2}{R^2U^4}$  for null geodesics (see \eq{vbigr})if we identify $l$ with $L$. Noting that for asymptotic spacetimes, we have $\frac{r}{2A}\left(\frac{B^\prime}{B}-\frac{A^\prime}{A}\right)\sim\frac{1}{r^m}$ with $m\ge1$ and  $8\pi r^2(p_{\text{rad}}-\rho)\sim\frac{1}{r^2}$ when $r\to\infty$. Thus, as long as $l$ is big enough, we always have $V_{2l}\approx V$. In section \ref{lrs}, we have found that $c^2=K$ is a critical point for the effective potential of null geodescis. For $c^2<K$, the effective potential $V$ has a local minimum, corresponding to a stable LR. So for gravitational perturbations, the effective potential $V_{2l}$ also has a local minimun when $l$ is large enough in the eikonal regime.
Therefore, we can infer that when $c^2<K$, i.e., a stable LR exists in the QBH spacetime, it becomes possible that the spectrum of linear QNMs contains the long-lived modes. Next, we are going to verify this relation by calculating the values of $\omega$.

Following standard numerical methods,  we would  use the method of direct integration to obtain the values of $\omega$. On the other hand, as shown in \cite{Festuccia:2008zx, Gurvitz:1988zz, Cardoso:2014sna}, when the potential  $V_{2l}$ has a local maximum and a local minimum (see \fig{diav}), for $l\gg1$, the real part of the frequency $\omega_R$ in four spacetime dimensions is given by the WKB approximation
\bea
\int_{r_a}^{r_b}\frac{dr}{\sqrt{B/A}}\sqrt{\omega_R^2-V_{2l}(r)}=\pi(n+1/2)=\int_{R_a}^{R_b}dRU(R)W(R)\sqrt{\omega_R^2-\bar{V}_{2l}(R)}\frac{dr}{dR}
\eea
where $n$ is a positive integer and $r_a$ and $r_b$ are two smaller roots of the equation $\omega_R^2-V_{sl}=0$ (see \fig{diav}). Obviously, we can easily conclude $\omega_R\in(V_{2l}^{min}, V_{2l}^{max})$. In addition, $R_a$ and $R_b$ are the roots corresponding to $r_a$ and $r_b$ in the $R$ coordinate, that is, $r_{a, b}=R_{a,b}U(R_{a,b})$. In addition, the imaginary part of the frequency $\omega_I$ is given by
\bea
\omega_I=-\frac{1}{8\omega_R \gamma}e^{-\Gamma}\,,
\eea
with
\bea
\Gamma&=&2\int_{r_b}^{r_c}\frac{dr}{\sqrt{B/A}}\sqrt{V_{sl}(r)-\omega_R^2}=\int_{R_b}^{R_c}dRU(R)W(R)\sqrt{\bar{V}_{2l}(R)-\omega_R^2}\frac{dr}{dR}\\
\gamma&=&\int_{r_a}^{r_b}\frac{dr}{\sqrt{B/A}}\frac{\cos^2\chi(r)}{\sqrt{\omega_R^2-V_{sl}(r)}}=\int_{R_a}^{R_b}dR\frac{U(R)W(R)\cos^2\bar{\chi}(R)}{\sqrt{\omega_R^2-\bar{V}_{2l}(r)}}\frac{dr}{dR}\\
\chi(r)&=&-\frac{\pi}{4}+\int_{r}^{r_b}\frac{dr}{\sqrt{B/A}}\sqrt{\omega_R^2-V_{2l}(r)}\non
&=&\bar{\chi}(R)=-\frac{\pi}{4}+\int_{R}^{R_a}dRU(R)W(R)
\sqrt{\omega_R^2-\bar{V}_{2l}(R)}\frac{dr}{dR} \,,
\eea
where $r_c=R_cU(R_c)$ is the largest root of the equation $\omega_R^2-V_{sl}=0$. We show the values of $\omega$ in \fig{qnmw} from both the numerical integration and WKB method. From the figure, we can see that the results given  by the two methods agree very well. Furthermore, the imaginary parts of the results show that the QNMs are indeed long-lived modes.

On the other hand, we would like to give some comments on the case $c^2\ge K$, in which there have been found no stable LRs in section \ref{lrs}. As shown in the \fig{vplot}, the effective potential $V_{2l}$ would have no local maximum in the eikonal limit. Following the similar analysis for planar AdS black holes in \cite{Festuccia:2008zx}, one can see that there are no long-lived modes survived in the linear perturbations of gravitations when $c^2\ge K$, that is, there are no stable LRs.

\section{Conclusion}
We have revisited the quasi-black hole spacetimes. In the parameter range $0<c^2<2/27$, we have found two LRs of which the inner one is stable and the outer one is unstable. They disappear altogether for $c^2>2/27$ with the critical value $c^2=2/27$ corresponding to a degenerate LR.  It's worth noting that near the black hole limit, i.e., $c\to 0$, we found there always exists a stable inner  LR while it has no correspondence in the black hole solution (when $c$ is strictly zero). We have also calculated the quasinormal modes of the QBHs. The WKB  method and numerical method both suggest that long-live modes survive when light rings exist, indicating that quasi-black hole spacetimes containing LRs, as a kind of ultracompact objects, may not be stable. Compared to the previous results on ultracompact stars that the exteriors are  Schwarzschild solutions, the LRs in the QBH model can turn on smoothly at a critical value of $c$. Therefore,  from another perspective, our work provides a concrete example to support the conjecture that observation of light rings may be strong evidence for black holes.

The quasi-black hole model indicates that LRs could be a signature exclusive for black holes.
However, a quasi-black hole cannot be treated as a real astrophysical body, since it has the total charge $q$ equal to the mass $M$, while in astrophysically reasonable situations, the charge is usually much smaller than the mass \cite{waldbook}. Thus, future studies shall focus on more realistic objects. For example, it would be important to consider UCO models with rotation since astrophysical bodies with high spin-mass ratio have been observed.

% In addition, from the results in \cite{Cunha:2017qtt, Guo:2020qwk}, we know that %LRs for UCOs always come in pairs. It would be interesting to explore UCOs %containing more than two LRs to verify whether the conjecture is still true.

\section*{Acknowledgments}
The work is in part supported by NSFC Grant   No. 11775022 and 11873044.

%\section*{References}
%\bibliographystyle{utphys}
%\bibliography{qnm}

\end{document}